\documentclass[article]{IEEEtran}
\IEEEoverridecommandlockouts
\advance\topmargin-1in\advance\oddsidemargin-1in
\evensidemargin\oddsidemargin
\usepackage{geometry}
\makeatletter

\def\@normalsize{\@setsize\normalsize{12pt}\xpt\@xpt
\abovedisplayskip 10pt plus2pt minus5pt\belowdisplayskip \abovedisplayskip
\abovedisplayshortskip \z@ plus3pt\belowdisplayshortskip 6pt plus3pt
minus3pt\let\@listi\@listI}

\def\thesection{\arabic{section}}

\renewcommand\thesubsection{\thesection.\arabic{subsection}}

\makeatother

\usepackage{cite}
\usepackage{amsmath,amssymb,amsfonts}
\usepackage{algorithmic}
\usepackage{graphicx}
\usepackage{textcomp}
\usepackage{xcolor}
\usepackage{amsmath}
\usepackage{wrapfig}
\usepackage{upgreek}
\usepackage{caption}
\usepackage{subcaption}
\usepackage{tabularray}
\usepackage{pifont}% http://ctan.org/pkg/pifont
\usepackage{color}
\usepackage{threeparttable}
\definecolor{DustyGray}{rgb}{0.588,0.588,0.588}
\renewcommand{\figurename}{\small\bfseries Figure}
\renewcommand{\tablename}{\small\bfseries Table}

\begin{document}

%%%%%%Heading%%%%%%
\title{\bfseries\LARGE{\textcolor{black}{Scalable Superconductor Neuron with Ternary Synaptic Connections for Ultra-Fast SNN Hardware}}}

\author{\IEEEauthorblockN{Mustafa Altay Karamuftuoglu, Beyza Zeynep Ucpinar, Arash Fayyazi,\\ Sasan Razmkhah, Mehdi Kamal, Massoud Pedram \vspace{-1.1em}}

\thanks{This work has been funded by the National Science Foundation (NSF) under the project Expedition: (Design and Integration of Superconducting Computation for Ventures beyond Exascale Realization) project with grant number 2124453. (Corresponding author: M. A. Karamuftuoglu)\\
M.A.K., B.Z.U., A.F., S.R., M.K., and M.P. are with Ming Hsieh Department of Electrical and Computer Engineering, University of Southern California, Los Angeles, USA. (\{karamuft, ucpinar, fayyazi, razmkhah, mehdi.kamal, pedram\}@usc.edu)
}
}

%%%%%%Abstract%%%%%%
\maketitle
\begin{abstract}
A novel high-fan-in differential superconductor neuron structure designed for ultra-high-performance Spiking Neural Network (SNN) accelerators is presented. Utilizing a high-fan-in neuron structure allows us to design SNN accelerators with more synaptic connections, enhancing the overall network capabilities. The proposed neuron design is based on superconductor electronics fabric, incorporating multiple superconducting loops, each with two Josephson Junctions. This arrangement enables each input data branch to have positive and negative inductive coupling, supporting excitatory and inhibitory synaptic data. Compatibility with synaptic devices and thresholding operation is achieved using a single flux quantum (SFQ) pulse-based logic style. The neuron design, along with ternary synaptic connections, forms the foundation for a superconductor-based SNN inference.
To demonstrate the capabilities of our design, we train the SNN using snnTorch, augmenting the PyTorch framework. After pruning, the demonstrated SNN inference achieves an impressive 96.1\% accuracy on MNIST images.
Notably, the network exhibits a remarkable throughput of \textcolor{black}{8.92} GHz while consuming only \textcolor{black}{1.5 nJ} per inference, including the energy consumption associated with cooling to 4K.
These results underscore the potential of superconductor electronics in developing high-performance and ultra-energy-efficient neural network accelerator architectures. 
\end{abstract}

\vspace{-1.0em}
%%%%%%Abstract%%%%%%

\section{Introduction}

Neuromorphic computing aims to revolutionize information processing by emulating the functionality of the biological brain. Within this field, spiking neural networks (SNN) are considered a promising energy-efficient solution for neuromorphic computing due to their sparse neural activity based on spike events. Although SNNs are more power-efficient than artificial neural network (ANN) methods, achieving high-performance metrics in power and speed with conventional hardware can be challenging. Therefore, exploration of new hardware paradigms is necessary since architectural manipulation alone is insufficient.

Superconductor logic has demonstrated remarkable improvements in operating speed and power consumption compared to conventional devices \cite{holmesISCNF2023}. Superconducting materials exhibit non-dissipative behavior below the critical temperature $T_c$, enabling energy-efficient data propagation at one-third the speed of light \cite{razmkhahBook}. In superconductor logic, Josephson Junctions (JJs) serve as ultra-fast switching elements, generating quantized SFQ voltage pulses \cite{likharev1991a}. These SFQ logic circuits operate at frequencies in the tens of GHz range and consume about $10^{-19}$ Joules per switch. The pulse-based information transfer and processing mechanisms in superconductor devices are akin to the characteristics of SNN components, making hardware implementations of superconductor neuromorphic devices a promising alternative. Various works have demonstrated different superconductor neurons \cite{hiroseNeuron2006, crottyNeuron2010, jardineNeuron2023, karamuftNeuron12023, karamuftNeuron22023, Friedman2024Neuron}.

Despite all the benefits, superconductors have some shortcomings when it comes to circuit implementation. One significant issue is the fan-in and fan-out of one. To bring an SFQ pulse to multiple inputs of the next stage, using pulse \textit{splitter} circuits becomes inevitable. Therefore, achieving neuronal characteristics with limited synaptic connections becomes crucial \cite{kenSegallFaninFanout}. In this work, we address these challenges and propose a superconductor neuron design that satisfies the dynamics of dendrites and a somatic operation, which functions as a threshold operation for activation. The proposed neuron device consists of multiple branches representing dendrites, with each branch placed between two JJs that determine the neuron's threshold value. The resistors on each branch establish the crucial aspects of the Leaky Integrate and Fire (LIF) neuron, achieving biologically plausible dynamics of asynchronous threshold operation.

Regardless of how well the superconductor neurons represent the functionality of their biological counterparts, their capabilities must be demonstrated in realistic implementations. Therefore, we employ the proposed neuron to construct a midsize inference feed-forward SNN. For network training, we utilize snnTorch, an extension of the PyTorch framework \cite{eshraghianSNNTORCH2021}. During training, we employ ternary synaptic values, allowing an SFQ pulse to carry out the information on the input data effectively. Ternary weights provide an intermediate state where a weight of 0 indicates disconnections between neurons that are useful where sparsity is desired. This approach could enhance memory utilization in a resource-constrained environment. 
Additionally, it is possible to connect the neuron dendrites to different synaptic devices to demonstrate the input weighting operation. To evaluate the network's performance, we use the MNIST dataset \cite{lecunMNIST1998} for training, achieving 97.07\% and 96.1\% accuracy pre- and post-pruning, respectively. Overall, the proposed neuron design enables the accurate and efficient realization of SNN implementations, making it suitable for various neuromorphic architectures.

The key contributions of this paper are as follows:
\begin{itemize}
\item Introducing a novel high-fan-in superconductor LIF neuron structure designed to create ultra-high-throughput and ultra-low-energy superconductor SNN accelerators.
\item Utilizing a neuron structure that involves only two JJs, independent of the dendrite size, resulting in an exceptionally low-power implementation.
\item Developing a neuron structure that supports positive and negative thresholds, enabling in-situ flexibility for various applications.
\item Demonstrating the effectiveness of the proposed neuron design by implementing an inference superconductor SNN for pattern recognition and verifying its computational behavior in a large-scale implementation.
\end{itemize}

\section{Background}
\subsection{Neuron model}
A biological neuron comprises dendrites that receive inputs from other neurons, a soma that performs the threshold operation, and an axon that transmits the output signal to succeeding neurons. In our superconductor neuron design, similar to the branching structure of neuron dendrites, we implement multiple superconducting loop branches. We must incorporate this structure into the analysis. The mathematical expression for an artificial neuron inspired by its biological counterpart is given in equation~\ref{eq:functionality}. However, our design involves multiple branches, and this equation needs modification to represent the operation accurately. In this modified equation, \textit{B} represents the number of branches, and \textit{I} represents the number of synaptic connections on each branch. For the dendrite size of 64, we have eight branches, each with four positive and four negative inputs. Increasing the dendrite count is as simple as adding a new branch to the neuron circuit.

Initially, input $x_{b, i}$ is multiplied by its synaptic weight $w_{b, i}$, where \textit{b} and \textit{i} correspond to the branch ID and input data port ID, respectively. A single branch corresponds to a group of input data ports. The expression $w^{1} \times x^{1}$ and $w^{2} \times x^{2}$ is written this way because a single port receives two weighted inputs in our hardware. Here, the \textit{w} values are adjustable coefficients defined by synaptic devices. Each multiplication result captures a post-synaptic potential transmitted to the cell body in a biological neuron from its dendrite.
After the multiplications, the results are combined and compared to a threshold value. If the sum $y$ exceeds the threshold $Th$, the output will be 1; otherwise, it will be 0. It is essential to consider each output port since our neuron supports both positive and negative thresholds. The output value determines the firing state of the neuron.

\begin{equation} \label{eq:functionality}
\begin{split}
y = \sum_{b=1}^{B} (\sum_{i=1}^{I} w_{b,i}^{1} \times x_{b,i}^{1} - w_{bi}^{2} \times x_{bi}^{2}) \\
(Output_{p}, Output_{n}) \left\{ 
\begin{array}{ c l }
    (1, 0) & \quad \textrm{if } y \geq Th_\mathrm{p} \\
    (0, 1) & \quad \textrm{if } y \leq Th_\mathrm{n} \\
    (0, 0) & \quad \textrm{otherwise}
\end{array} \right.
\end{split}
\end{equation}

\subsection{Single flux quantum pulse generation}
\begin{figure}[!t]
\centering
    \includegraphics[width=0.76\linewidth]{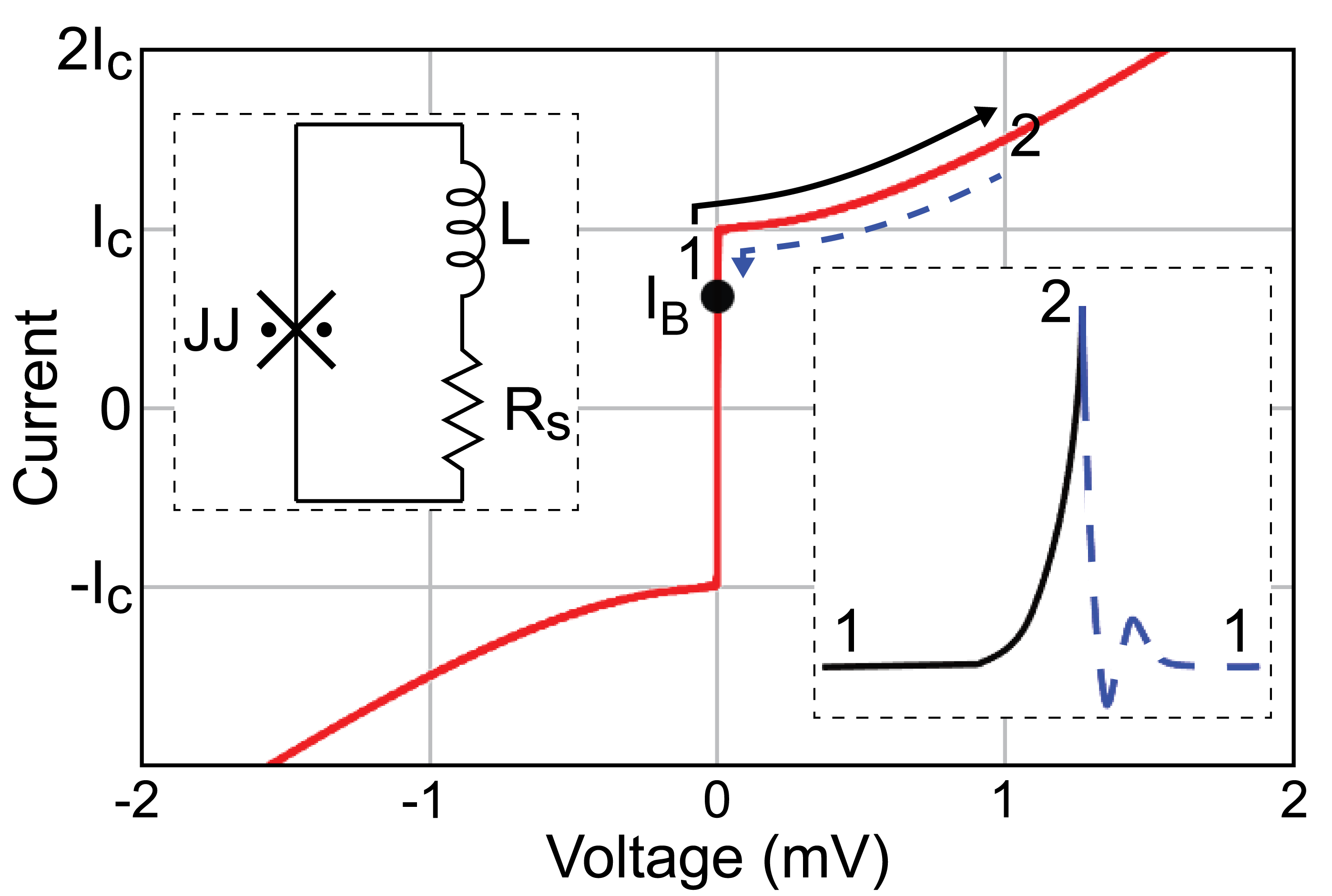}
    \caption{\textcolor{black}{I-V curve of a JJ and SFQ pulse generation.}}
    \label{fig:sfqPulse}
\end{figure}

\textcolor{black}{Josephson junction (JJ) is the key component in superconductor circuits that generates an SFQ pulse. JJ compromises two superconducting electrodes separated by a weak link, such as a thin insulating barrier. In SCE, the JJ is current bias to a value such as $I_{B}$, as illustrated in Fig. \ref{fig:sfqPulse}. At low temperatures, the supercurrent exhibits quantum behavior, allowing it to exist across JJ without any voltage.}

\textcolor{black}{Upon receiving an external current, the total current surpasses the critical current $I_{C}$ of JJ, triggering a transition into the resistive state from point 1 to point 2. As the JJ behaves like a resistive element, it produces a voltage pulse during this transition. Subsequently, the overall current reverts to its initial point 1, indicating the JJ's return to the superconducting state. As a result, the energy dissipated on this transition on the current-voltage (I-V) curve is equivalent to one flux quanta ($\Phi_{0}$). This entire process occurs in a few picoseconds, highlighting the generation of an SFQ pulse. In Fig. \ref{fig:sfqPulse}, $R_{s}$ corresponds to a shunt resistor to modify the I-V characteristics of the JJ, and L represents the parasitic inductance. The dynamic behavior of the JJ can be described by a set of equations, namely DC and AC Josephson equations as given in \cite{razmkhahBook}.}

\begin{table*}[!htbp]
\centering
\caption{The example neuronal devices with different models for CMOS and superconductor technologies.}
\label{table:neuronComparison}
\resizebox{0.98\textwidth}{!}{%
\begin{tabular}{|c|c|c|c|c|c|} 
\cline{2-6}
\multicolumn{1}{c|}{}  & \textbf{Chen et al \cite{chenNeuron2023} } & \textbf{Aamir et al \cite{aamirNeuron2018}} & \textbf{Jardine et al \cite{aamirNeuron2018}} & \textbf{Crotty et al \cite{crottyNeuron2010}} & \begin{tabular}[c]{@{}c@{}}\textbf{Proposed Neuron}\end{tabular} \\ \hline
\textbf{Technology} & CMOS (65 nm)  & CMOS (65 nm) & \begin{tabular}[c]{@{}c@{}}Superconductor\\(Hybrid RSFQ$^*$-QFP$^{**}$)\end{tabular} & Superconductor (RSFQ)  & Superconductor (RSFQ) \\ \hline
\textbf{Neuron Model}  & Leaky Integrate and Fire & \begin{tabular}[c]{@{}c@{}}Adaptive-exponential\\Integrate and Fire\end{tabular} & Integrate and Fire & Integrate and Fire & Leaky Integrate and Fire \\ \hline
\textbf{\textbf{Components}} & \begin{tabular}[c]{@{}c@{}}Leaky-integrator (a PMOS, an NMOS\\and a capacitor), input, and fire devices\\(each with a PMOS and an NMOS)\end{tabular} & \begin{tabular}[c]{@{}c@{}}Source-degenerated OTA,\\membrane capacitor and\\multiple pass transistors\\for different modes\end{tabular} & \begin{tabular}[c]{@{}c@{}}A coupled bias line and\\a two-JJ loop with coupled \\inductors (AQFP$^{***}$ buffer\\as an activation)\end{tabular} & \begin{tabular}[c]{@{}c@{}}Two-JJ loop\\with inductors\end{tabular} & \begin{tabular}[c]{@{}c@{}}Multiple two-JJ loop\\with resistors and\\coupled inductors\end{tabular}  \\ 
\hline
\textbf{Frequency} & Hz & MHz & GHz & GHz & GHz \\ \hline
\end{tabular}}
\begin{tablenotes}
      \small
      \item $^*$Rapid Single Flux Quantum, $^{**}$Quantum Flux Parametron, $^{***}$Adiabatic QFP
\end{tablenotes}
\end{table*}

\subsection{Related work}
In superconductor electronics, pulse storage and transmission are achieved using superconducting quantum interference device (SQUID) structures \cite{kleinerSQUID2004}. Read and write operations with SQUIDs are accomplished using superconductor electronic circuitry. Similarly, practical implementations of superconductor neuromorphic components also revolve around SQUID-like structures. Authors of \cite{hiroseNeuron2006, jardineNeuron2023} implement SQUIDs to induce excitatory and inhibitory effects within a neuron device by applying pulses in opposite directions. Toomey et al. \cite{toomeyNeuron2019} shows an artificial neuron with superconducting nanowires. However, the scalability of the design is an issue. Crotty et al. \cite{crottyNeuron2010} present a different configuration of two JJs placed in a superconductor loop as a neuron design. In contrast, the circuits shown at \cite{karamuftNeuron12023, karamuftNeuron22023} demonstrate a SQUID structure interrupted by a resistor as somatic operations for a neuron.

Example studies for different neuron devices with various technologies, including the proposed design, are presented in Table \ref{table:neuronComparison}. For superconductor devices, the neuron behavior is realized using a superconductor loop consisting of two Josephson Junctions (JJs). These designs perform integrate and fire operations upon receiving current from the input ports. The inclusion of resistor components introduces a leaky behavior in the loop. In the case of CMOS devices, the membrane potential is mimicked by adding a capacitor, and the overall designs revolve around different placements of NMOS and PMOS transistors. When comparing the frequency of operation, superconductor neurons outperform CMOS counterparts by three orders of magnitude. This significant speed advantage is attributed to the distinct switching elements, transistors in CMOS and JJs in superconductors, arising from the underlying technological differences between the two technologies.

\textcolor{black}{The studies in \cite{CALIF2023, SpikingYOLO2020} demonstrated the effective generation of both positive and negative output spikes through the use of positive and negative thresholds in a high-level model. The authors highlighted the practical feasibility of this dual output generation and underscored the inherent benefits of the spiking neuron model. Consequently, the integration of this strategy into deep neural networks may offer substantial improvements in accuracy.}

Regarding the background of superconductor neural networks, various papers explore the possibility of using superconductor devices for large-scale neuromorphic applications. Schneider et al. \cite{schneiderNetwork2017} demonstrate character recognition using a two-layer SNN with MJJs. Bozbey et al. \cite{bozbeyNetwork2020} utilize superconducting neurons and hybrid synaptic interconnects for inference SNN. Ishida et al. \cite{ishidaNetwork2021} present a processing element for neuromorphic applications. Cai et al. \cite{caiMNIST2019} present a DNN framework using AQFP superconducting technology, and Zhang et al. \cite{zhangNetwork2023} show the study on SNN with quantum phase-slip junction (QPSJ).

\section{Superconductor-based LIF Neuron}
\subsection{Design implementation}

\textcolor{black}{Development of neural networks using superconductor circuits requires the integration of specialized components designed for synaptic and somatic operations. Due to the limited density, memory, and fan-in/out attributes of the standard SFQ cells, incorporating CMOS-appropriate architectures like MAC (Multiply-Accumulate) would lead to higher hardware costs within the SFQ circuit design methodology. Therefore, opting for SNNs that incorporate neuromorphic components becomes the most practical hardware strategy for SFQ-based neural networks.}

In a standard SFQ library, such as RSFQ and Energy-efficient RSFQ (ERSFQ), various interconnect cells perform pulse-based operations, including Josephson transmission line (JTL), passive transmission line (PTL), splitter (SPL), and confluence buffer (CBU) \cite{likharev1991a, kirichenkoERSFQ2011}.
The JTL cell serves as a buffer, transmitting the input to the output with a picosecond delay, effectively mimicking the behavior of an axon. Conversely, the PTL is a waveguide that performs the same operation for longer distances while introducing a negligible delay. Notably, PTLs do not have any active elements, resulting in no power consumption, excluding the PTL driver and receiver circuits.
SPL cells increase the driveable fan-out count, representing the branching axon endings. CBU implements the functional operation of converting data from parallel to serial. It functions similarly to an asynchronous OR gate, generating a single output pulse when the input pulses arrive nearly simultaneously (within a "small" timing window).

\begin{figure*}[!t]
\centering
    \includegraphics[width=1\linewidth]{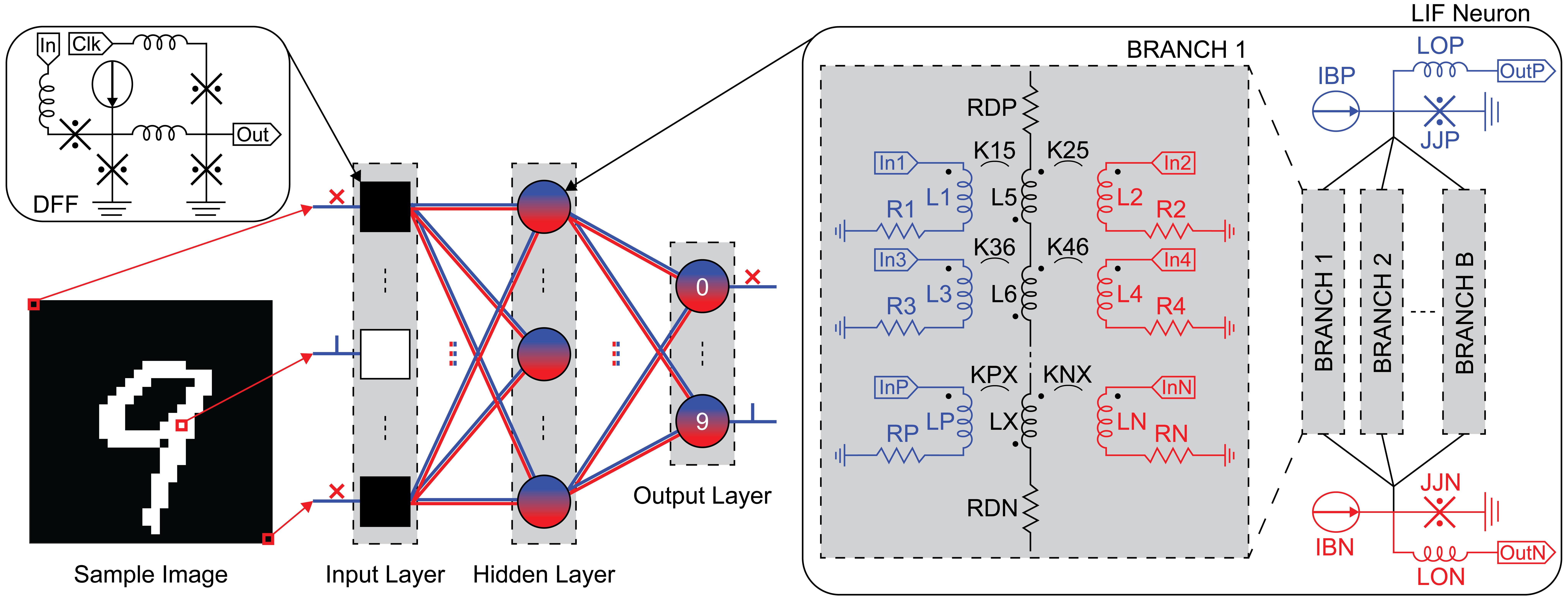}
    \caption{\textcolor{black}{Network architecture comprising a standard D Flip-Flop (DFF) and Leaky Integrate and Fire (LIF) neuron. DFFs provide input pulses to the network and perform an initial synchronization for the data flow. On the right side, a neuron circuit schematic includes multiple branches to accommodate the high-fan-in feature. In this configuration, the neuron performs the leaky accumulation within the resistive loops and activation function with JJs.}}
    \label{fig:neuronSchematic}
\end{figure*}

To achieve high-synaptic connections, neurons should support the simultaneous and multiple incoming pulses. Our designed superconductor neuron is compatible with the standard library cells while satisfying the abilities of a biological neuron, as depicted in Fig.~\ref{fig:neuronSchematic}. When comparing the superconductor neuron to its biological counterpart, the dendrites correspond to inductive couplings ($L1$ - $L5$) over a superconducting loop.
The positive and negative couplings, such as $K15$ and $K25$, determine the excitatory and inhibitory effects on the aggregated current between the Josephson Junctions $JJP$ and $JJN$. As the current directions for excitatory and inhibitory inputs are opposite, the remaining current in the loop represents the subtraction of the injected current from the couplings. The positive sign of the remaining current corresponds to the supercurrent circulating from $JJN$ to $JJP$, and vice versa for the negative sign.
To introduce the leaky behavior, resistors RDP and RDN are incorporated into the structure. Their values also set the time constant $\tau$ for the neuron decay.

To increase the dendrite count, new branches can be introduced by adding $RDN$-$LXN$-...-$LXP$-$RDP$ with their inductive couplings between $JJN$ and $JJP$ without affecting the power consumption of the neuron. However, the maximum number of neuron dendrites depends on the employed fabrication technology. Another approach is to include new inductive couplings in the existing branches. However, it's important to note that the overall inductance of the branches determines the time constant for the neuron. Including new inductive couplings in existing branches will increase this time constant if each inductor value is not decreased. Therefore, after establishing a branch with four positive and four negative inputs, we opted to add new branches to the simulation structure.

In our design, we have two bias ports on top of the Josephson Junctions (JJs), denoted as $IBP$ and $IBN$. These bias ports introduce an offset for the inputs, resulting in a lower threshold value. By adjusting the bias currents $IBP$ and $IBN$, we can tune the threshold values $Th_p$ and $Th_n$ to desired levels. Increasing the bias current results in the neuron triggering with lower incoming energy, hence reducing the threshold.
In the schematic shown in Fig. \ref{fig:neuronSchematic}, we have added JTL buffers to each output port for observing the pulses in our testbench. The design's propagation delay is in the order of picoseconds, providing a fast response time.

In this work, the inputs from the couplings contribute to the current stored in the superconducting loop. As our inputs are Single Flux Quantum (SFQ) pulses, the \textit{x} input in Eq. \ref{eq:functionality} will be logic '0' or '1', depending on the absence or presence of the pulse, respectively.
Furthermore, the synaptic weights are represented as ternary values, with \textit{w} assigned as '-1', '0', or '1'. Before the input ports, any desired synaptic device can be utilized to modify the applied input value. For instance, a magnetic Josephson Junction (MJJ) can behave like a locally adjustable synaptic element and can be employed as an interconnect \cite{schneiderMagneticJJ}. Another example is a hybrid implementation of CMOS and superconductor components, as presented in \cite{bozbeyNetwork2020, razmkhah2023hybrid}, for the SNN inference.

\subsection{Design flexibility}
Depending on the application, the couplings in the superconducting loops can be interchangeably utilized. For example, it is possible to use only the positive output side of the neuron while connecting the input ports to any desired synaptic device, including memory and sensor arrays. Simultaneously, the negative output can sink to the ground to exclude its contribution.
Alternatively, it is possible to utilize only excitatory inputs and exclude inhibitory effects altogether. In this scenario, the two output sides of the neuron can be used for decisions when the available synaptic weights are limited. For instance, even when using only positive weights, it is still possible to achieve inhibitory effects by utilizing the pulse generated from the negative side of the neuron. Both positive and negative couplings were established to evaluate their impact on the network. However, in our large-scale implementations, we utilized the neuron in an inference SNN where only the positive output port was used.  

\textcolor{black}{For the given design, the threshold plays a crucial role in determining the neuron's firing state. 
Typically, the threshold for SFQ-based neurons is set during the design phase and remains fixed after the tape-out. The solutions to overcome this challenge introduce circuit complexity. However, the threshold can be modified during runtime by leveraging input ports in the neuron structure. In such instances, the threshold is determined based on specific inputs serving as effective offset, influencing the neuron's response. This threshold adjustment mechanism underscores the flexibility and adaptability inherent in the networks.}

\section{Superconductor-based SNN implementation}
To support our preference for SNNs in neuromorphic applications, we compare MAC-based computational units for ANNs to spike-based synapses and neuron units of SNNs using SFQ circuits.
In MAC-based DNN (Deep Neural Network) implementations, a separate memory unit is required, which leads to slower processing as weight values need to be fetched from memory for each data value. Additionally, computations are performed on shared MAC units in a time-multiplexing manner, making them computationally slower. To mitigate this, many parallel MAC units can be employed, resulting in a bulky design due to high JJ counts and latency.
On the other hand, SNNs achieve weighting by utilizing local synaptic devices between neurons, which corresponds to a single multiplication operation compared to the multi-stage operations in MAC units. This eliminates the delay between data flow from memory to computation units, making SNNs more area and power-efficient. However, they may achieve slightly lower inference accuracy compared to ANNs.

%%%%%%
\textcolor{black}{
In the case of the MAC-based neuronal computation unit, SFQ-based implementations of adders and multipliers could be extracted by synthesizing their HDL model using the qPALACE tool, the ColdFlux's Design Compiler \cite{Coldflux2023}. 
Our studies show that the 4-bit multiplier has 16 stages and uses 3200 JJs running at 17.5 GHz clock frequency.
Additionally, the 8-bit Kogge-Stone adder (KSA) has eight stages and includes 2786 JJs running at up to 19.1 GHz.
Considering the high JJ counts and latency due to the multi-stage operations in the MAC units of ANNs, SFQ-based SNNs can be more area- and power-efficient by utilizing the parallel computation of spike-based components.}

\textcolor{black}{The burst of pulses for the synaptic designs such as an NDRO with multi-flux storing characteristics \cite{mndroZeynep2023} may introduce more delay due to the time interval between the pulses and the current limitations on superconductor fabrication technologies for MJJs can be challenging to achieve consistent properties across a network. Therefore, we suggest employing ternary synaptic values to achieve high throughput and power efficiency due to lower neuronal activity. This also affects the decay time of the proposed neuron since there will be a single pulse rather than an arbitrary number of pulses.}

In a feed-forward SNN architecture shown in Fig. \ref{fig:neuronSchematic}, input spikes propagate through synaptic elements for each data flow, and the activation function is performed by neuron elements that receive weighted inputs from synapses. In our proposed neuron, the synaptic connection is achieved by simply connecting the input with a PTL receiver since we utilize ternary synaptic values. It is also possible to employ single magnetic Josephson Junctions (MJJ) on synaptic connections to support other weight coefficients. However, MJJ fabrication is still ongoing research, and scalability is problematic. Even though our demonstration is for a feed-forward architecture, the proposed neuron can be utilized in different SNN architectures.

\subsection{Data quantization}

\begin{wrapfigure}{iR}{0.23\textwidth}
    \begin{center}%\centering
    \includegraphics[width=0.23\textwidth]{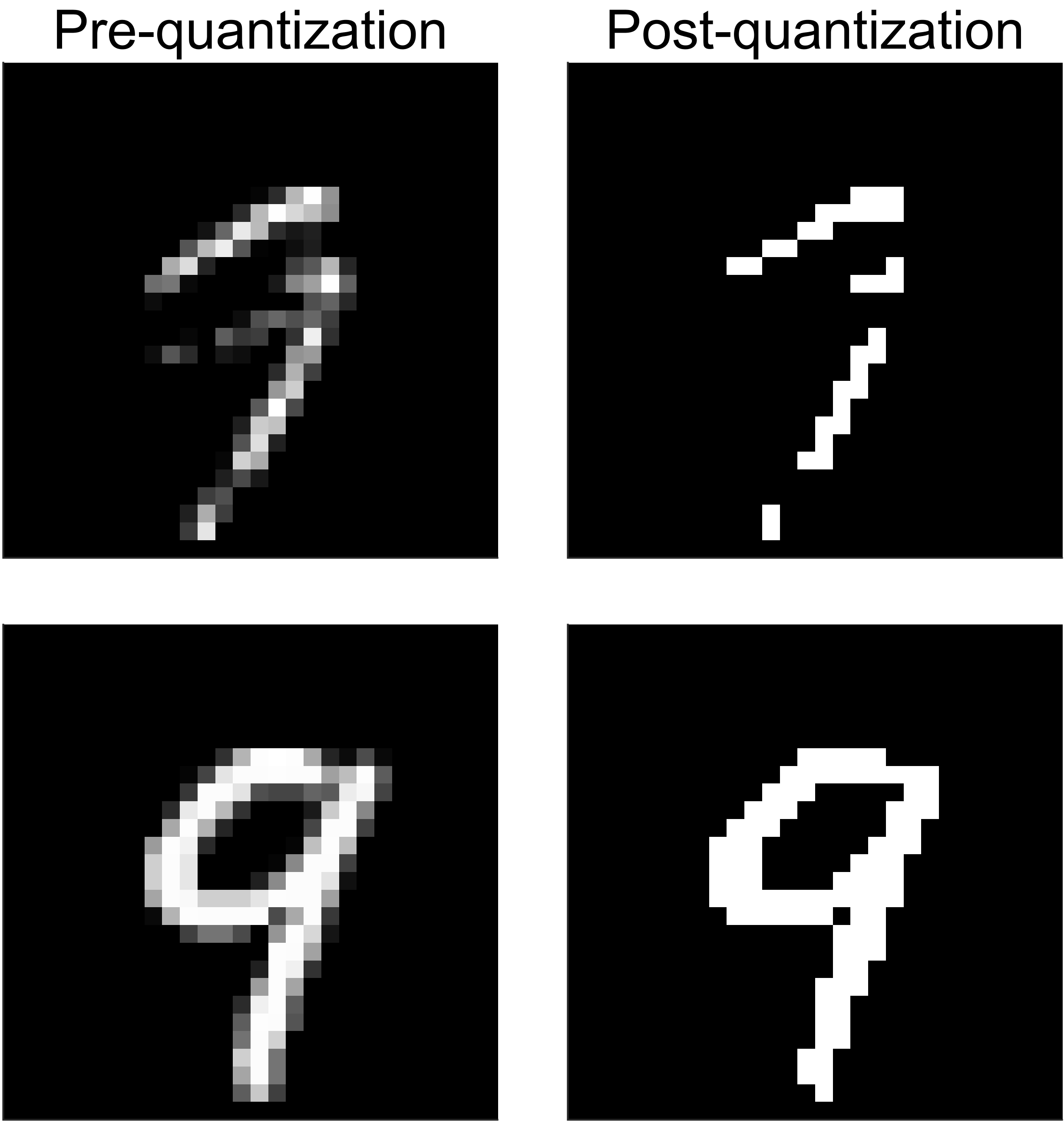}
    \end{center}
    \caption{Two different pre- and post-quantization images for the number 9. Each image is 28$\times$28 pixels.}
    \label{fig:quantization}
\end{wrapfigure}

In the inference phase, we employ 1-bit quantization for input data. This extreme form of quantization represents the data with the presence or absence of an SFQ pulse. The weights are limited to values of '-1', '0', and '1', resulting in the input current from each data to the neuron being the same for all input ports but with possibly opposite directions.
This quantization strategy reduces memory requirements and enables faster computation in a highly parallel manner, albeit at the cost of lower accuracy in the final classification. The sample images for pre- and post-quantization are shown in Fig. \ref{fig:quantization}, where white pixels on the post-quantization image correspond to SFQ pulses during the computation. Throughout this process, the pixels colored in gray are rounded to the nearest integer, representing either black or white.

Although this quantization may lead to the loss of essential aspects of the data, as seen in the example on the top row, there are situations where it may have little impact, as demonstrated in the bottom row example.
It is important to note that even though we use 1-bit resolution for data representation when providing input to the network, applying the Poisson distribution allows us to achieve non-deterministic behavior, introducing randomness to the pixel data during training \cite{heegerPoisson2000}.

\subsection{Training and pruning}
During the training process, snnTorch leverages gradient-based optimization techniques, where continuous approximations of gradients are used for error propagation. The network receives information with spiking patterns, introducing different timings and orders of spikes. To enhance the training process, the number of steps is defined for the data, supplying inputs in a non-deterministic way using the Poisson distribution. Additionally, the decay time and overall behavior of neurons are defined to match the characteristics of the underlying hardware.

To reduce the network size, the network could be pruned by considering the maximum $k$ synaptic connections for each neuron. Thus, in the training process, for each neuron, we retain the $k$ weights with the most significant absolute values once a mini-batch is entirely processed. The remaining weights are masked, effectively setting them to zero for the forward pass of the subsequent mini-batch. Only the unmasked weights are updated using their corresponding gradients after completing the forward pass for a mini-batch. This process may lead to a different set of weights being masked by the end of the mini-batch.

As we approach the end of the training process, the mask may undergo minor changes. After completing the training process, we replace the masked weights with zeros and keep the unmasked weights at their final values. The weight vector of each neuron, which is fixed and sparse, is then utilized during inference.
Dead neurons and their corresponding synaptic connections will be removed by network pruning, reducing computational costs and improving energy efficiency. We maintain the maximum dendrite count of the neuron in the network, enabling the utilization of the same computational module in a time-multiplexed manner. However, this approach will decrease the throughput as it requires maintaining intermediate computation results over multiple iterations. Nevertheless, it remains area-efficient in producing the final result.

\section{Simulation Results}
\subsection{Single neuron demonstration}
\begin{figure}[!t]
\centering
    \includegraphics[width=1\linewidth]{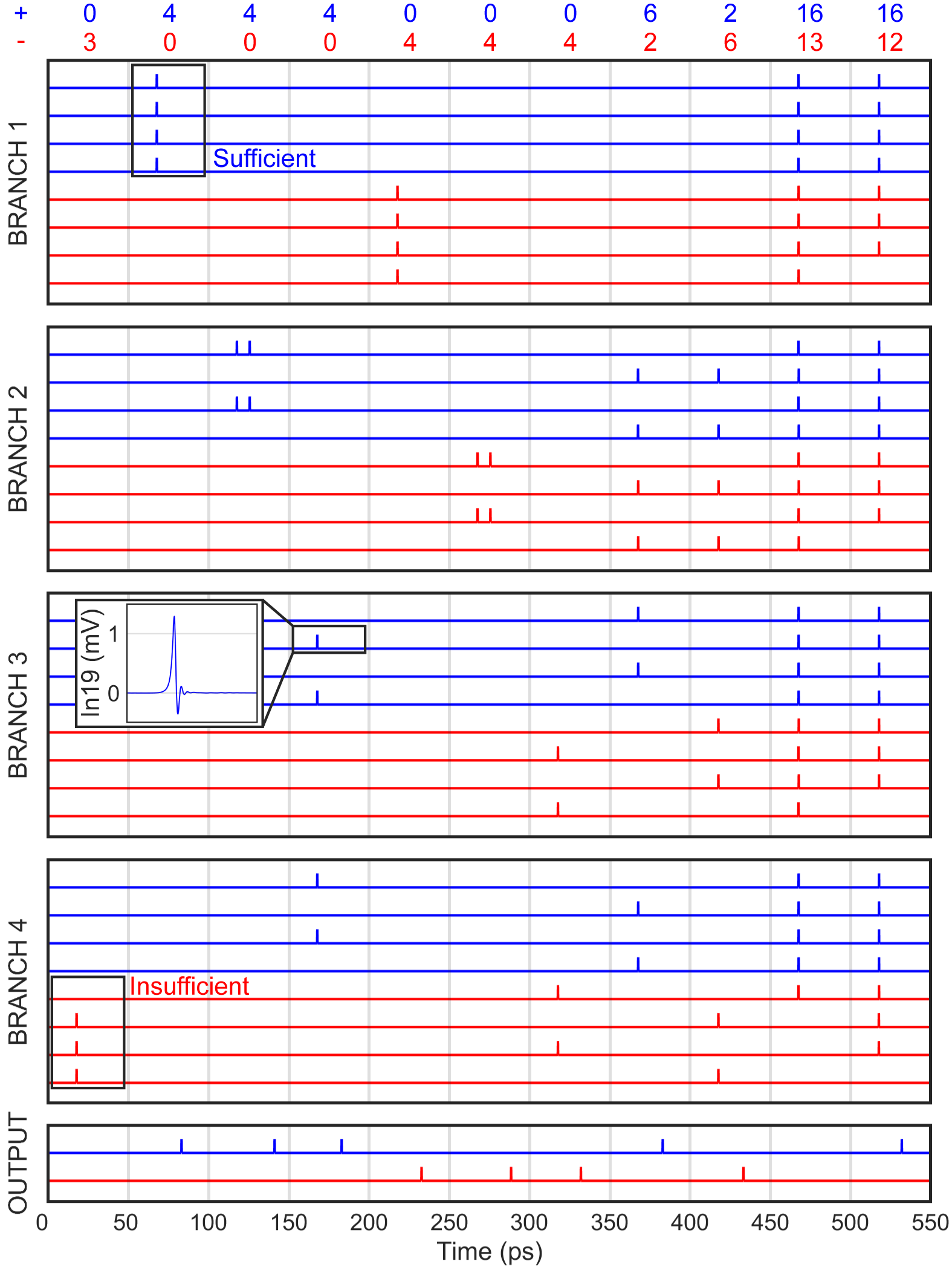}
    \caption{The simulation of the proposed neuron with 32 synaptic connections was performed using JoSIM, where the threshold and input frequency were set to 4 SFQ pulses and 20 GHz, respectively. The simulation results are presented in the form of plots displaying the converted JJ phase data for better visualization. In these plots, the observed pulses from the JJs in JTLs connected to the neuron are evident. The color blue represents positive inputs, while negative inputs are displayed in red.}
    \label{fig:neuronSimulation}
\end{figure}

The neuron is a critical component of the network, and verifying its correct functionality is of utmost importance.
For the analog simulation, we designed a superconducting neuron with 32 synaptic connections, consisting of 16 excitatory and 16 inhibitory connections. The neuron has four branches, each with four positive and four negative inductive couplings between the Josephson Junctions (JJs). 
\textcolor{black}{In the simulations, the properties of our JJs were modeled based on the SFQ5ee fabrication process developed for superconductor electronics at MIT Lincoln Laboratory \cite{MITLL_tolpygo2015}. The critical component of the neuron circuit is the JJ, and the neuron behavior strongly depends on the employed JJ features. 
The neuron threshold is mainly determined by the critical current of the JJ and its relative bias current. The bias for superconductor circuits is externally applied with high precision, enabling accurate control for the correct functionality.} 
In this design, all coupling values are the same since the injected current from the inputs is expected to be the same for an SFQ pulse on each input port. The bias and JJ values are configured to set the threshold for receiving 4 SFQ pulses for both the positive and negative sides.
To perform the analog simulations, we used the SPICE-based Josephson simulator (JoSIM) \cite{delportJoSIM}, and the results are reported in Fig. \ref{fig:neuronSimulation}. The simulations confirm the correct functionality of the designed superconducting neuron, demonstrating its ability to integrate excitatory and inhibitory inputs and perform threshold operations as expected.

In the simulation, we initially provided 3 SFQ pulses from the negative coupled pins. Since the threshold is set at 4 SFQ pulses, no pulse is observed at the negative output. Next, we provided 4 SFQ pulses, and an SFQ pulse was observed at the positive output, indicating the successful integration of excitatory inputs and threshold operation.
Furthermore, we demonstrated the capability of applying arbitrary pulses between the 100 ps and 150 ps time instances in Fig. \ref{fig:neuronSimulation}. The output pulse was observed with a short delay from the last arriving pulse, confirming the correct functionality of the neuron.
To verify the correct operation for inhibitory inputs, we repeated the previous patterns but applied them to the negative side. The simulations confirmed the correct operation of the neuron for both excitatory and inhibitory inputs.
In a corner case scenario, we tested the neuron with 16 excitatory and 12 inhibitory inputs just after the 500 ps time instance in Fig. \ref{fig:neuronSimulation}. In this case, if one additional inhibitory input arrives at the neuron, the threshold value cannot be reached, as shown right before the 500 ps time instance.
For this simulation, the $I_\mathrm{c}$ values for JJN and JJP are designed as 100 $\mu$A, while each JJ bias current from IBN and IBP is set to approximately 92 $\mu$A, contributing to the correct functionality of the neuron.

\begin{figure}[!t]
\centering
    \includegraphics[width=1\linewidth]{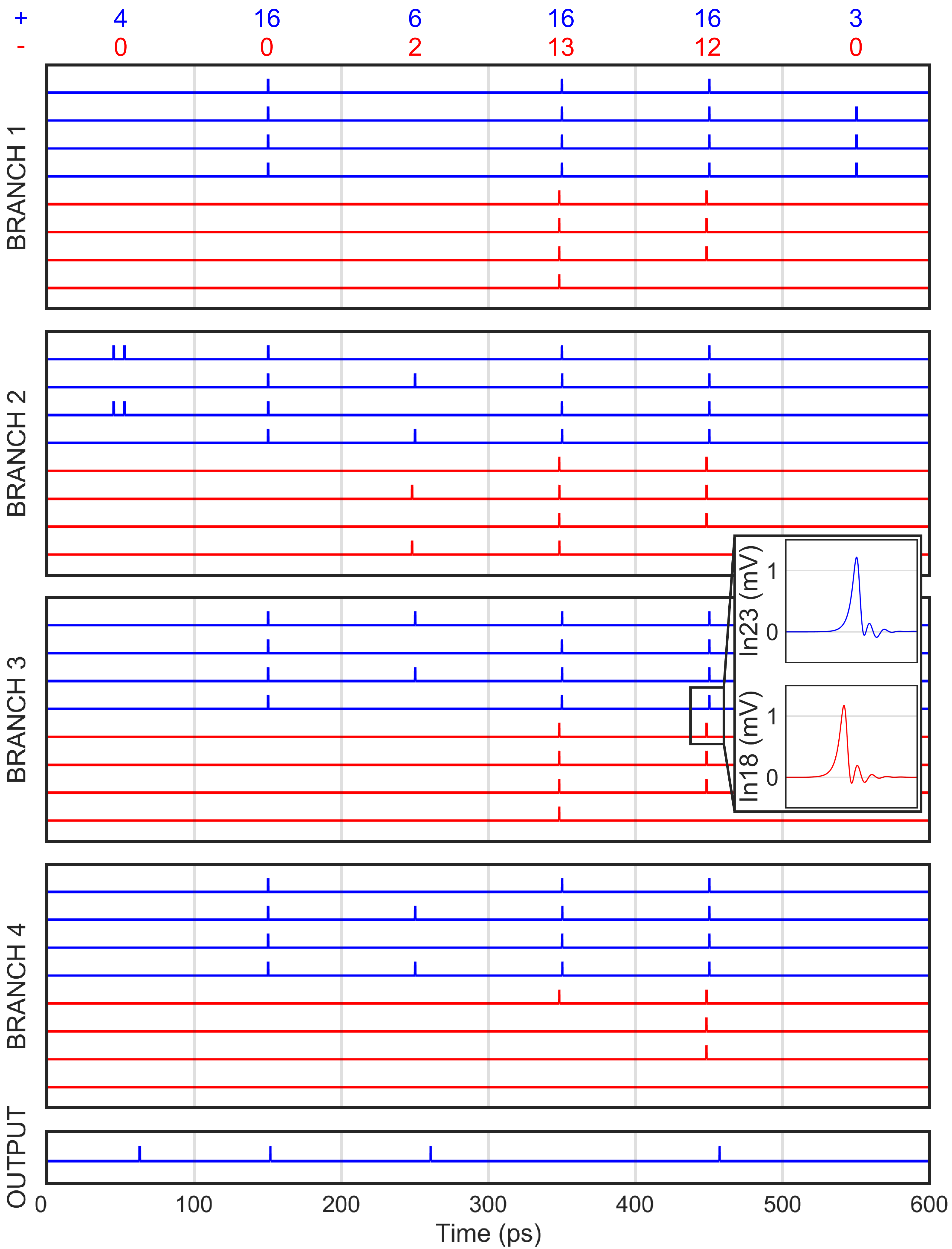}
    \caption{\textcolor{black}{The simulation result of the proposed neuron where negative pulses arrive earlier than the positive pulses for the initialization. The neuron threshold and input frequency were set to 4 SFQ pulses and 10 GHz, respectively. The JJ phase data is color-coded and converted for better visualization. \small(LP = 6.0pH, LX = 11.63pH, LN = 4.8pH, LOP = 0.2pH, JJP = $82\mu A$, RP = RN = 0.37 $\Omega$, RDP = 3.36 $\Omega$, RBP = 32.1 $\Omega$, KPX = 0.43, KNX = -0.59)}}
    \label{fig:neuronSimulation2}
\end{figure}

\textcolor{black}{In the context of neural network design, a conventional approach involves leveraging the positive output side \textit{JJP} of the neuron while excluding the negative side \textit{JJN}, as illustrated in Fig. \ref{fig:neuronSchematic}. However, this standard configuration introduces a significant challenge associated with the timing of positive and negative inputs due to the asynchronous decision-making processes of neurons. The temporal misalignment between positive and negative pulses risks the accuracy of the system, particularly when positive spikes arrive earlier than the negative counterparts. This temporal discrepancy may result in the junction \textit{JJP} switching early and generating erroneous outputs. Therefore, understanding and addressing the input timing concerns may improve the reliability and precision of the network under various operational conditions.}

\textcolor{black}{To address this challenge, a strategic solution involves providing negative pulses earlier than positive ones and establishing a negative offset as an initialization. This approach mitigates issues related to the delay of negative pulses introduced by jitter. However, due to the leaky behavior, the impact of negative pulses is diminished when positive pulses arrive. Consequently, rather than utilizing a symmetrical structure for positive and negative fan-in ports, we increase the effect of negative inputs by increasing the mutual coupling \textit{KNX}. The simulation result of this approach is shown in Fig. \ref{fig:neuronSimulation2}.}

\textcolor{black}{The simulation presents multiple cases to validate the correct functionality of the proposed neuron. To meet the required timing for the traces of negative pulses, the input frequency for the test patterns decreased to 10 GHz, and the maximum frequency for this setup is observed as 10.75 GHz. In the first 100 ps, multiple pulses are received from a single branch (BRANCH 2). Subsequently, we tested a corner case where all positive input ports received an SFQ pulse. In this case, only a single output pulse is generated due to the refractory time of the neuron that is determined by the \textit{JJP}. Next, the example cases with positive and negative pulses are tested. Lastly, we applied three SFQ pulses in a scenario that is insufficient to surpass the threshold of 4 pulses.}

\subsection{Neural network demonstration}

\begin{table*}[t]
\centering
\caption{The comparison of MNIST results for different studies.}
% \caption{\small\bfseries The comparison of MNIST results for different studies.}
\label{table:allMNIST}
\resizebox{0.98\textwidth}{!}{%
\begin{tabular}{|c|c|c|c|c|c|} 
\cline{2-6}
\multicolumn{1}{c|}{}                                                         & \textbf{Bang et al \cite{bangMNIST2021}}                                              & \textbf{Asghar et al \cite{asgharNetwork2023}}                                                  & \textbf{Cai et al \cite{caiMNIST2019}}                                                                         & \textbf{Zhang et al \cite{zhangNetwork2023}}                                                                                            & \textbf{This paper}                                                    \\ 
\hline
\textbf{Technology}                                                           & CMOS (65 nm)                                                          & CMOS (65 nm)                                                          & Superconductor (AQFP)                                                                                     & Superconductor (QPSJ)                                                                                          & Superconductor (RSFQ)                                                  \\ 
\hline
\textbf{Network Type}                                                         & SNN                                                                   & SNN                                                                   & DNN                                                                                              & SNN                                                                                                            & SNN                                                                    \\ 
\hline
\textbf{Architecture}                                                         & (28$\times$28)-200-200-10                                      & (14$\times$14)-30-20-10                                             & \begin{tabular}[c]{@{}c@{}}Conv3–Conv3–Pool–...\\Conv5–Conv5–Pool–...\\Conv7–FC500–FC800-Outlayer\end{tabular}                                                                               & (28$\times$28)-100-10 & (28$\times$28)-128-96-96-10                                             \\ 
\hline
\textbf{\# of synaptic weights}                                               & 198 k                                                               & 6.7 k                                                                 & 4 M                                                                                          & 79.4 k                                                                                                      & 21.1 k                                                                   \\ 
\hline
\textbf{Accuracy (\%)}                                                        & 97.46                                                                  & 96.56                                                                 & 96.95                                                                                            & 86.1                                                                                                          & 96.1                                                                  \\ 
\hline
\textbf{Throughput (Hz)}                                                      & N/A                                                               & 2.03 M                                                                   & 6.67 M                                                                                           & N/A                                                                                                         & \textcolor{black}{8.92 G}                                                                   \\ 
\hline
\textbf{Energy(J)/inference}                                                  & 1.18 $\mu$                                                            & 261 n                                                            & 111 f                                                                                             & N/A                                                                                                         & \textcolor{black}{1.5 n}                                                            \\
\hline
\end{tabular}}
\vspace{-1em}
\end{table*}

% AQFP Energy per inference: 0.37 p x300 = 111

In our high-level experiments, we constructed a fully connected SNN inference with three hidden layers and achieved an accuracy of 97.07\% on the MNIST dataset. The network design supports \textcolor{black}{8.92} GHz throughput that is based on the decay time of the neuron and interconnects created on each layer as a pipeline stage. It is worth noting that the number of layers does not impact the throughput of giga inferences per second; however, it does contribute to higher inference delay due to the pipeline architecture demonstration. Additionally, we did not apply any augmentation techniques to the training or test data in this work.

The input layer consists of 784 buffers, and the neuron counts for the three hidden layers are 128, 96, and 96, respectively. For the fanout, the SPL and PTL driver/receiver cells introduce additional delays. The maximum fanout for the network is observed to be 88, which corresponds to 5 stages in an SPL tree, with each SPL supporting a fanout of three. In our library, each SPL and the combination of PTL driver/receiver cells approximately introduce a 3.1 ps and a 3.6 ps delay, resulting in a total delay of 15.5 ps from 5 SPL stages and a 3.6 ps from the data propagation. As a consequence, the maximum delay of a pre-synaptic pulse from a neuron becomes 19.1 ps.

The thresholds for the neurons, representing the number of SFQ pulses required before an action potential is generated, are set to 4, 2, and 2 for the hidden layers and 2 for the output layer, respectively. These threshold values have been determined through experimental analysis.
During the training process, snnTorch generates synaptic values for the entire network. The training is performed over five steps with a total of 300 epochs.
Since the initial training process does not consider any neuron's dendrite size, we perform network pruning to enhance network performance while maintaining accuracy. For practicality, we prune the network to have a maximum of 64 synaptic connections per neuron and subsequently eliminate dead neurons along with their synaptic connections.

As a result of pruning, the overall synaptic count is reduced from 122304 to 21120, leading to an 82.73\% reduction in static power for the connections. However, there is a 0.97\% accuracy loss associated with these savings. Additionally, the fanout of 109 of the input pixels is reduced to zero, indicating that 675 of the input pixels are sufficient to determine most of the classifications.
% JTL   +   SPL  +  Neu  +  PTLRX  +  PTLTX  +  DFF  + Neu Load JTL
% 1575*JTL + 15853*SPL + 330*Neu + 21120*TX + 21120*RX + 675*DFF  + 330*JTL
% 1575*2   + 15853*4   + 330*1   + 21120*2  + 21120*4  + 675*4   + 330*2 = 196972
% JJ 196972
Considering the fanout of 3 for the SPL, the pruned network will contain 196,972 JJs, even though the neurons only use 330 JJs. In this calculation, we also include the input provided by DFFs in the input layer and JTLs for delay balancing in the data paths. \textcolor{black}{Moreover, due to our approach, where positive pulses are applied with a delay, additional JTLs are required. Since half of the remaining synapses are positive, 10,560 additional JTLs where a single additional JTL is placed on every positive dendrite of a neuron are required to provide delayed positive pulses into the neurons. As a result, the total JJ count becomes 207,532.}

\textcolor{black}{The power consumption of each PTL driver, PTL receiver, JTL, SPL with a fanout of 3, and DFF in the library is measured at 0.663$\mu$W, 0.492$\mu$W, 0.449$\mu$W, 1.225$\mu$W, and 0.533$\mu$W, respectively.
Based on these values, the total static power consumption of the whole network, including the transmission lines, is estimated to be 44.6 mW. As a result, the energy consumption per inference is calculated as 5.01 pJ.
Additionally, when considering the energy consumption associated with cooling to 4K, the overall consumption increases to 1.5 nJ per inference.}

In Table \ref{table:allMNIST}, we provide an application-level hardware implementation comparison of different neuromorphic designs using the MNIST dataset. Our SNN inference, built using the proposed neuron, achieves an accuracy above 96\% through pruning, supervised learning, and ternary synaptic values.
While our inference accuracy may not be as high as that of the CMOS counterpart, the superconductor-based designs exhibit significantly higher inference throughput. It is worth noting that the AQFP design does not have static power consumption, but our design improves energy efficiency compared to the CMOS-based design. Furthermore, existing literature shows that it is possible to implement resistor-free SNNs using ERSFQ by replacing resistors with shunted JJs \cite{kirichenkoERSFQ2011}, thus eliminating static power consumption and achieving much higher energy efficiency compared to CMOS and approaching AQFP levels.

\section{Conclusion}
In this work, we present a hardware implementation of a high-fan-in superconductor neuron structure. The analog simulation of the circuit is demonstrated using JoSIM, showcasing the functionality and behavior of the neuron. We further showcase the capabilities of the neuron in SNN inference using the popular MNIST dataset. The fully connected network achieves an impressive 97.07\% accuracy. To improve power efficiency and reduce static power consumption, we utilize pruning techniques on the network, resulting in a slightly reduced accuracy of 96.1\%. Despite this accuracy reduction, the pruned network retains the ability to process data at the order of GHz throughput. The inference network is trained using snnTorch, which leverages gradient-based optimization techniques for continuous approximations of the gradients during training. Significantly, the neuron operates asynchronously, allowing the overall flow to proceed without the need for a clock signal, except for the input buffers on the input layer, which are used to synchronize the data flow.

\section{Acknowledgment}
\textcolor{black}{The authors would like to thank Armin Abdollahi (USC) and Mahdi Nazemi (USC) for their help in the configuration of the neural network framework.}

\bibliographystyle{IEEEtran}
\bibliography{references}
\end{document}